# Edge Computing for Microgrid via MATLAB Embedded Coder and Low-Cost Smart Meters

1st Linna Xu
School of System Science and Engineering
Sun-Yat Sen University
Guangzhou, China
xuln6@mail2.sysu.edu.cn

2nd Jian Huang
School of System Science and Engineering
Sun-Yat Sen University
Guangzhou, China
huangj629@mail2.sysu.edu.cn

3rd Shan Yang
School of System Science and Engineering
Sun-Yat Sen University
Guangzhou, China
yangsh237@mail2.sysu.edu.cn

Yongli Zhu*
School of System Science and Engineering
Sun-Yat Sen University
Guangzhou, China
yzhu16@alum.utk.edu
corresponding author

*Abstract*—In this paper, an edge computing-based machine-learning study is conducted for solar inverter power forecasting and droop control in a remote microgrid. The machine learning models and control algorithms are directly deployed on an edge-computing device (a smart meter-concentrator) in the microgrid rather than on a cloud server at the far-end control center, reducing the communication time the inverters need to wait. Experimental results on an ARM-based smart meter board demonstrate the feasibility and correctness of the proposed approach by comparing against the results on the desktop PC.

## I. INTRODUCTION

Microgrids have become an important part of the modern power grid because of their high efficiency, flexibility, and environmental friendliness. One of the most frequently used renewable sources in a microrod is the photovoltaic (PV) generation. Forecasting techniques aim at accurately predicting PV power generation, and accurate PV power forecasts can effectively reduce the impact of PV power fluctuations on the microgrid. In [1], the E**X**treme **G**radient **B**oosting (XGBoost) model is used for solar PV power forecasting and has achieved good performance. In [2], a graph neural network method is adopted to predict the solar PV power in multiple PV farms. In [3], a support vector machine (SVM)-based approach is developed to improve the forecasting accuracy for PV power output. In [4], the artificial neural network (ANN) is leveraged to capture the nonlinear dependencies between the weather information and the PV power output.

Besides the research efforts in PV power forecasting, another prevailing research direction in the microgrid area is the droop control strategy [5][6], which adaptively adjusts PV inverters' (active and/or reactive) power outputs to satisfy a specific goal (e.g., maintaining a secured voltage level) by using a set of elegant, analytic formula.

The usual practice to deploy the above machine learning models or control algorithms is using the cloud service at a central server; however, this cannot suit the needs for 1) low latency (the round trip to the cloud server can cost extra time) 2) enhanced privacy (the data communication with the cloud server can expose the data to potential attack) and 3) affordable cost (hosting algorithms or models on high-performance commercial cloud can be expensive).

Therefore, a diversity in deployment targets has emerged, with more and more devices running locally near the end-users rather than on a central cloud server for security, reliability, and performance considerations. For example, the previous centralized solution might not be applicable for a remote microgrid with weak, low-bandwidth communication infrastructures (e.g., microgrids in rural mountains or islands). Thus, advances in modern low-cost embedded hardware can be taken advantage of, i.e., the models or algorithms can be directly deployed on an edge device in the microgrid (e.g., a smart meter-concentrator at the PCC bus of the microgrid).

However, devising and deploying an edge computing (on-device computing) architecture for the microgrid is not a simple task: firstly, the original algorithm or machine learning models have to be converted or cross-compiled towards various target devices; secondly, given the limited CPU capability and memory size of the low-cost edge device, the converted code is expected to run at an acceptable speed. Therefore, the cross-compiled code must be optimized toward the target device's specific hardware and software architecture.

This paper presents a study about edge computing for a remote microgrid using the MATLAB Embedded Coder. Two example cases are considered and implemented on a real meter-concentrator board with ARM CPU: 1) PV inverters' power forecasting and 2) PV inverters' droop control strategy.

In the remaining part of this paper, Section II describes the PV inverters, the microgrid, and the two edge-computing use cases considered in this paper. Section III presents the experiment results of the machine learning models (for inverter power forecasting) and droop controllers (for inverter voltage regulation) on a desktop PC. Section IV illustrates the basic workflow of the MATLAB Embedded Coder and demonstrates the deployment results on a smart meter board. Conclusion and future work are given in the final section.

## II. TWO EDGE COMPUTING CASES FOR A REMOTE MICROGRID

### A. Power Forecasting for PV Inverters

In a microgrid, solar PV is typically integrated at or near the end-user side via a DC/AC inverter (usually less than 1MW), as shown in Fig .1. In the first use case, we want to respectively forecast the PV inverters' active and reactive power output given the terminal bus measurements (three-phase voltage magnitude, three-phase current magnitude, previous power setting point, power factors, etc.).

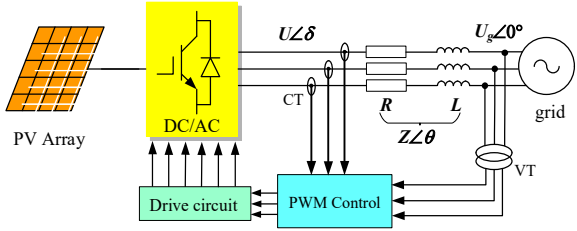

Fig. 1. The diagram of a typical grid-tied solar PV inverter

Note that this task is not identical to the task of *forecasting solar PV generation* that emphasizes the maximum possible PV array's power generation under a given weather condition (e.g., irradiation, temperature); our task here uses only ordinary electricity measurements. The motivation behind this use case is that not every microgrid or inverter is equipped with weather measurement devices. Moreover, since the inverter is directly interfaced with the grid, its power output is more useful in higher-level control applications. The adopted machine learning framework in this paper is shown in Fig. 2.

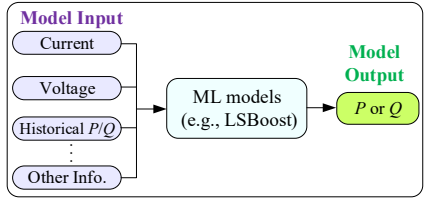

Fig. 2. The machine learning framework for PV inverter power forecasting.

### B. Droop Control Strategy for PV Inverters

The second use case is about the inverter's droop control strategy. For a given inverter shown in Fig. 1, its active and reactive power can be described by the following formula [7]:

$$P = \left(\frac{U_g U}{Z}\cos\delta - \frac{U_g^2}{Z}\right)\cos\theta + \frac{U_g U}{Z}\sin\delta\sin\theta$$
$$Q = \left(\frac{U_g U}{Z}\cos\delta - \frac{U_g^2}{Z}\right)\sin\theta - \frac{U_g U}{Z}\sin\delta\cos\theta \quad (1)$$

where $Z=|R+jX|$ and $\theta$ are respectively the line impedance magnitude and impedance angle. $\delta$ is the power angle. When the line impedance is weak resistive (i.e., $Z \approx X$), the above formula can be approximated respectively by :

$$P = \frac{U_g U}{X}\delta, \quad Q = \frac{U_g(U-U_g)}{X} \quad (2)$$

Then, the following droop control scheme is adopted in this paper for each inverter's voltage regulation:

$$Q_{ref} = Q_{min} + k_q(U_{max} - U_{meas}), \quad Q_{min} \leq Q_{ref} \leq Q_{max}$$
$$k_q = (Q_{max} - Q_{min})/(U_{max} - U_{min}) > 0 \quad (3)$$
$$P_{ref} = \sqrt{S_{rate}^2 - Q_{ref}^2}, \quad 0 \leq P_{ref} \leq S_{rate}$$

where $S_{rate}$, $Q_{min}$, and $Q_{max}$ are respectively the inverter's MVA capacity, lower and upper limits of the reactive power. $U_{min}$, $U_{max}$ are respectively the (allowable) minimum and maximum voltage levels, e.g., 0.9pu (198V) and 1.1pu (242V) in this paper. $k_q$ is the droop-control coefficient. $P_{ref}$ and $Q_{ref}$ are the setting points calculated for each inverter, which will be sent (by the PCC bus concentrator) to all the downstream inverters.

### C. Dataset from A Real Microgrid

In this paper, we obtain real measurement data (voltage, current, power, etc.) of four solar PV inverters from a real microgrid in a remote village, as illustrated in Fig. 3.

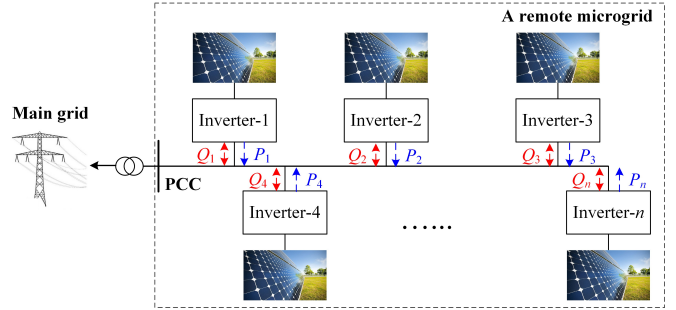

Fig. 3. An illustration of the remote microgrid with solar PV inverters.

Regarding machine learning for PV inverter power forecasting, the dataset was collected for about 30 days (from May to June 2024). The dataset consists of three-phase currents and voltages, active and reactive power at a 15-minute resolution. Preprocessing, such as missing-value-filling and dirty-data-cleaning, is applied to the raw dataset to improve the data quality. The data was split into a training set and a test set with a ratio of 80%:20%.

Regarding the implementation of the droop control strategy, power capacity parameters ($S_{rate}$, $Q_{min}$, and $Q_{max}$) are collected for all the inverters of that microgrid.

Both the trained machine models and droop control algorithms will be deployed on our edge-computing device (i.e., an ARM-based smart meter) (Section IV) and tested against the results obtained on the PC (Section III).

## III. MODELS AND ALGORITHMS DEVELOPED ON PC

In this section, all the experiments are implemented on a desktop PC with AMD Ryzen 2.1GHz CPU and 16GB RAM.

### A. ML Model Development for Inverter Power Forecasting

Here, LSBoost is chosen as the main method, a variant of decision-tree methods that utilizes ensemble learning techniques by combining multiple simple decision trees. It also employs the idea of "boosting" by iterative creating new trees to fit the older trees' residuals.

The prediction label is the actual active/reactive powers. The feature input of the machine learning model is all the data other than the active/reactive power. The machine learning models are trained using LSBoost (least square boosting tree) based on the field measurements of the PV inverters and compared with SVM and ANN.

The LSBoost is essentially an additive model, which trains $K$ trees on totally $n$ data samples and then gives prediction for the $i$-th data sample by Eq. (4):

$$\hat{y}_i = \sum_{k=1}^{K} f_k(x_i) \tag{4}$$

where $f_k$ represents one sub-decision tree, each sub-tree is trained by progressively optimizing an associated objective function. The gradient boosting algorithm gradually minimizes the residuals at each step. The objective function at the *m*-th step is shown in Eq. (5).

$$Obj^{(m)} = \sum_{i=1}^{n} l(y_i, \hat{y}_i^{(m-1)} + f_m(x_i)) + \Omega(f_m) \tag{5}$$

The above objective function has two parts: the loss function term *l* and the regularization term Ω. A highly performant regression model can be obtained by minimizing such objective functions and continuously adding new trees.

The performance metrics used in this study are $R^2$ and MAPE. $R^2$ is an index between 0 and 1, as shown in Eq. (6). It measures the explanatory power of the independent variables over the dependent variable, the larger the better.

$$R^2 = 1 - \frac{\sum_{i=1}^{n}(y_i - \hat{y}_i)^2}{\sum_{i=1}^{n}(y_i - \bar{y})^2} \tag{6}$$

MAPE indicates the percentage of forecasting errors relative to the true values. Since the measured power can have zero values, the following formula is leveraged to prevent the "divide-by-zero" issue (*Cap* is the inverter's power capacity).

$$MAPE = (1 - \sqrt{\frac{1}{n}\sum_{i=1}^{n}(\frac{y_i - \hat{y}_i}{Cap})^2}) \times 100\% \tag{7}$$

### B. PV Inverter's Reactive Power Forecasting

We utilized the MATLAB platform to train three models: LSBoost, SVM, and ANN based on the reactive power data from six photovoltaic inverters. The comparative results for the three models are shown in Tables I and II.

TABLE I. $R^2$ ON REACTIVE POWER TEST SET

| Inverter No. | LSBoost | SVM | ANN |
|---|---|---|---|
| 1 | **0.9626** | 0.8416 | 0.8451 |
| 2 | **0.9939** | 0.9541 | 0.9697 |
| 3 | **0.9999** | 0.9837 | 0.9995 |
| 4 | **0.9993** | 0.4025 | -0.0007 |

TABLE II. MAPE ON REACTIVE POWER TEST SET

| Inverter No. | LSBoost | SVM | ANN |
|---|---|---|---|
| 1 | **2.4027%** | 4.9457% | 4.8918% |
| 2 | **1.0393%** | 2.8395% | 2.3077% |
| 3 | **0.1095%** | 1.3759% | 0.2421% |
| 4 | **0.4025%** | 0.0885% | 14.9371% |

Tables Ⅰ and Ⅱ show that LSBoost outperforms SVM and ANN in terms of larger $R^2$ across most inverters. Notably, on inverter-3 and inverter-6, LSBoost respectively achieves 0.9999 and 0.9993 $R^2$ values, indicating near-perfect fits. In contrast, the performance of other models was significantly inferior to LSBoost. LSBoost also demonstrates significantly lower MAPE values than SVM and ANN across all inverters. For example, on inverter 3, LSBoost had a MAPE of 0.1095%, which is substantially lower than SVM's 1.3759% and ANN's 0.2421%.

In addition, we plot the reactive power predictions by different models for the PV inverter-2 across a complete day, as shown in Fig. 4.

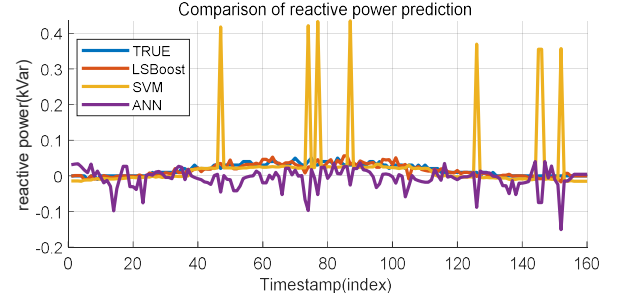

Fig. 4. Comparison of reactive power prediction for PV inverter-2

### C. PV Inverter's Active Power Forecasting

Similarly, we train active power prediction models for all four PV inverters. The comparative results are presented in Tables III and IV.

TABLE III. $R^2$ ON ACTIVE POWER TEST SET

| Inverter No. | LSBoost | SVM | ANN |
|---|---|---|---|
| 1 | **0.9878** | 0.1329 | 0.9652 |
| 2 | **0.9981** | 0.3896 | 0.9664 |
| 3 | **0.9955** | -0.2738 | 0.9568 |
| 4 | **0.9987** | 0.9980 | 0.9979 |

TABLE IV. MAPE ON ACTIVE POWER TEST SET

| Inverter No. | LSBoost | SVM | ANN |
|---|---|---|---|
| 1 | **2.6110%** | 15.9253% | 4.4081% |
| 2 | **1.7760%** | 27.4764% | 7.4138% |
| 3 | **1.7547%** | 20.6670% | 5.4180% |
| 4 | **0.6897%** | 0.8378% | 0.8766% |

LSBoost again achieved higher $R^2$ than SVM and ANN across all inverters. For example, on inverter 6, LSBoost's MAPE was 0.6897%, notably lower than SVM's 0.8378% and ANN's 0.8766%. A plot of the active power predictions by different models for the PV inverter-2 in one complete day is shown in Fig. 5.

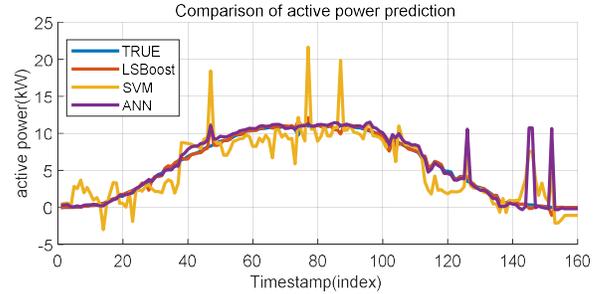

Fig. 5. Comparison of active power prediction for the inverter-2

## D. V-Q Droop Control for Inverter Voltage Regulation

Here, the *V-Q* droop control discussed in Section II.B is implemented on the PC. The calculated $P_{ref}$ and $Q_{ref}$ are shown in Table V for a series of fictitious voltage measurements.

TABLE V.    THE *P/Q* SETTING POINTS BY THE *V-Q* DROOP CONTROL

| Inverter No. | $P_{ref}$ (kW) | $Q_{ref}$ (kVar) | $k_q$ (kVar/V) |
|---|---|---|---|
| 1 | 15.0000 | 0.0000 | 0.3592 |
| 2 | 21.9317 | -12.0000 | 0.5986 |
| 3 | 18.7283 | -9.5000 | 0.5028 |
| 4 | 14.7792 | 6.1298 | 0.3831 |

## IV. ALGORITHM DEPLOYMENT ON EDGE DEVICE

In this section, the previously developed LSBoost models and droop-control algorithm will be deployed on an edge device, i.e., the ARM board of a smart meter.

### A. Description of the Edge Device

Fig. 6 displays the edge device where the model is deployed. The board in this study is based on ARMv8 architecture, having four Cortex-A processors and 1GB of disk space. The installed operating system is ARM Linux (version 4.9.38).

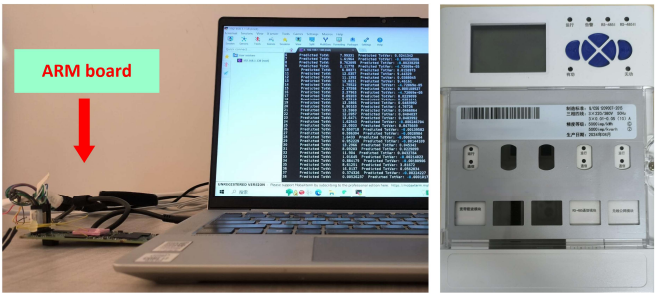

Fig. 6.   The ARM-based smart meter board used in this study.

### B. Introduction to Matlab Coder and Embedded Coder

*1) Matlab Coder:* Matlab Coder [8] is a code converter that can transform MATLAB code to C or C++ source code for various hardware platforms without manually re-coding efforts. It supports most of MATLAB's built-in functions and toolboxes. Besides, the Matlab Coder will implicitly do a lot of "code-optimization" work (e.g., memory alignment, function Inlining), trying to make the obtained C/C++ code run fast. Fig 7 and 8 respectively illustrate the workflow and GUI of using MATLAB Coder for source code generation.

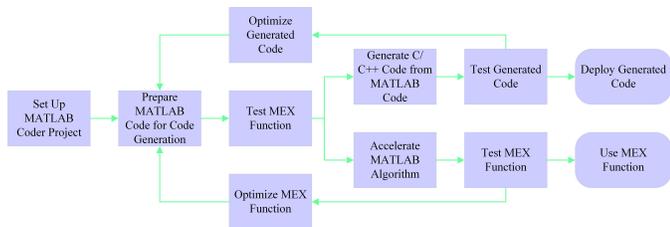

Fig. 7.   Code Generation Workflow of the MATLAB Coder.

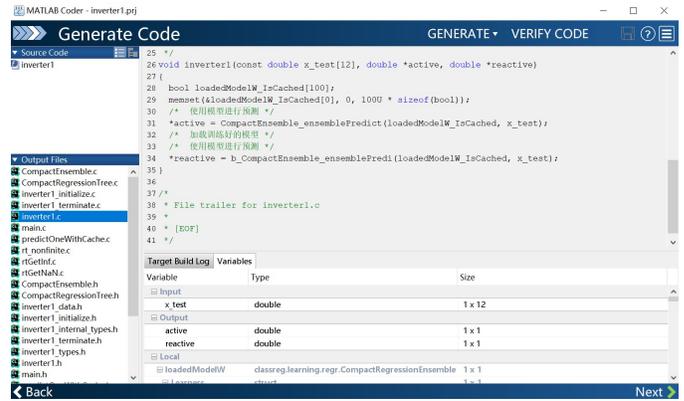

Fig. 8.   Source code generated by MATLAB Coder (GUI).

*2) Embedded Coder:* Embedded Coder [9][10] extends MATLAB Coder capacity with advanced optimization for the generated function, making the generated code more specific for the embedded environment. These extra optimizations promote code efficiency and facilitate integration with data types, calibration parameters, and legacy code. Fig 9 shows the code generation workflow involving the Embedded Coder. With the assistance of MATLAB Coder and Embedded Coder, we can convert the MATLAB script to C/C++ source code and then use gcc to cross-compile the source code to obtain the final executable file for the target device.

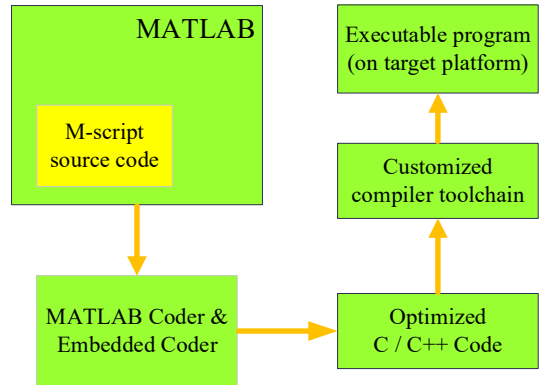

Fig. 9.   The code generation workflow involving the Embedded Coder.

### C. Deploy the Droop Control Algorithm on the Smart Meter

The deployment result of the V-Q droop control algorithm on the smart meter board is shown in Fig. 10, which is the same as the result in Table V.

```
/run# tftp -g -b 512 -l /run/voltageDroopControl
/run# ls
udev                    voltageDroopControl
/run# chmod a+x voltageDroopControl
/run# ls
udev                    voltageDroopControl
/run# ./voltageDroopControl
Pref:                   Qref:
 15.00000                0.00000
 21.93171              -12.00000
 18.72832               -9.50000
 14.77921                6.12983
```

Fig. 10. Result of the droop control algorithm on the smart meter board

## D. Deploy PV Reactive Power Forecasting Models on the Smart Meter

In this part, we present the deployment results of the PV reactive power forecasting models and compare them with the previous results obtained on PC.

```
double CompactRegressionTree_predict(const double obj_CutPredictorIndex[21],
  const double obj_Children[42], const double obj_CutPoint[21], const bool
  obj_NanCutPoints[21], const double obj_NodeMean[21], const double Xin[12])
{
  int m;
  bool exitg1;
  m = 0;
  exitg1 = false;
  while (!(exitg1 || (obj_CutPredictorIndex[m] == 0.0))) {
    double d;
    d = Xin[(int)obj_CutPredictorIndex[m] - 1];
    if (rtIsNaN(d) || obj_NanCutPoints[m]) {
      exitg1 = true;
    } else if (d < obj_CutPoint[m]) {
      m = (int)obj_Children[m << 1] - 1;
    } else {
      m = (int)obj_Children[(m << 1) + 1] - 1;
    }
  }

  return obj_NodeMean[m];
}
```

Fig. 11. Code snippets from the generated C/C++ code.

Fig. 11 illustrates some code snippets converted by the Embedded Coder from the well-trained LSBoost model in Section III. These C/C++ codes are then executed on the ARM board of the smart meter. As an example, Table VI lists a portion of predictions yielded by inverter-1's reactive power forecasting model (running respectively on PC and smart meter). From the results, it can be observed that the two outputs are basically the same.

TABLE VI. OUTPUTS OF INVERTER-1'S REACTIVE POWER MODEL ON THE TESTING SET

| PC | Smart Meter |
|---|---|
| 4.718668801 | **4.718669** |
| 4.709738436 | **4.709738** |
| 4.838473868 | **4.838474** |
| 0.039732713 | **0.039733** |

The identity of the PC's and smart meter's outputs are then checked by computing the MAPE and RMSE between the two. In Table VII, they are identical up to about six decimal points, which demonstrates that the machine-learning models are successfully deployed to the edge device with correct outputs.

TABLE VII. PREDICTION RESULTS COMPARISON: PC VS. SMART METER

| Inverter No. | MAPE | RMSE |
|---|---|---|
| 1 | 0.0197269383542% | 0.000000296542139 |
| 2 | 0.0730401175678% | 0.000000286189546 |
| 3 | 0.0492650314837% | 0.000000279593972 |
| 4 | 0.0091767734564% | 0.000000288723648 |

## E. Deploy PV Active Power Forecasting Models on the Smart Meter

Here, the comparison results of the active power control model on the PC and smart meter are present. Because the models for active power forecasting and reactive power forecasting both use LSBoost and possess the same type of input and output, the generated C/C++ source code and the deployment process are also similar.

TABLE VIII. OUTPUTS OF INVERTER-1'S ACTIVE POWER MODEL ON THE TESTING SET

| PC | Smart Meter |
|---|---|
| 10.58910111 | **10.589101** |
| 7.949180966 | **7.949181** |
| 2.494983789 | **2.494984** |
| 13.79637164 | **13.796372** |

TABLE IX. PREDICTION RESULTS COMPARISON: PC VS. SMART METER

| Inverter No. | MAPE | RMSE |
|---|---|---|
| 1 | 0.0002631108228% | 0.000000290281346 |
| 2 | 0.0014664632808% | 0.000000287272756 |
| 3 | 0.0002828037630% | 0.000000281836386 |
| 4 | 0.0007273273043% | 0.000000300903832 |

Table VIII shows a portion of predictions made on the PC and the smart meter, and Table IX shows their differences in terms of MAPE and RMSE. The results indicate that the deployed model for active power forecasting is also successful.

## F. Speed Comparison: PC vs. Smart Meter

Since the droop control algorithm is simple (c.f. Eq. (3)), the time differences between the program running on the PC and the smart meter are not remarkable. Thus, the speed comparison study is mainly conducted for the machine-learning use case.

Table X lists the average inference time costs for one input entry (executed respectively on the PC and the smart meter). The trained model runs slower on the smart meter (around 1ms) than on the PC since the desktop PC's CPU is much faster than the low-cost ARM processor. However, this speed performance is good enough for quasi-steady-state applications like voltage regulation or economic dispatch.

TABLE X. COMPARISON OF INFERENCE TIME: PC VS. SMART METER

| Inverter No. | Reactive power model | | Active power model | |
|---|---|---|---|---|
| | PC | Smart Meter | PC | Smart Meter |
| 1 | 0.1325ms | **0.9499ms** | 0.4233ms | **0.9779ms** |
| 2 | 0.1575ms | **0.9499ms** | 0.4180ms | **0.9724ms** |
| 3 | 0.1161ms | **0.9544ms** | 0.3826ms | **0.9862ms** |
| 4 | 0.2350ms | **0.9722ms** | 0.5465ms | **0.9775ms** |

## V. Conclusion

This paper shares two edge-computing use cases about using MATLAB Embedded Coder in a real microgrid project. Comparison results on the desktop PC and the smart meter-concentrator board demonstrate that the deployed model can perform the inference task in milliseconds, and the yielded results can match the results on the PC. Investigating other advanced use cases (e.g., deploying deep learning-based control algorithms on the smart meter) by the proposed edge-computing scheme can be the next step in this research line.